\documentclass[runningheads,a4paper]{llncs}
\usepackage{tabularx}
\usepackage{array}
\usepackage{booktabs}
\usepackage{float}
\usepackage{placeins}
\usepackage{graphicx}
\usepackage{placeins}
\usepackage{amsmath}
\usepackage{amssymb}
\usepackage{xcolor}
\usepackage{url}
\newcolumntype{L}[1]{>{\raggedright\arraybackslash}p{#1}}

\begin{document}
\raggedbottom

\mainmatter

\title{Ask the Peers: Ranking NFR Coverage Gaps for Requirements Review}


\titlerunning{Ranking NFR Coverage Gaps for Requirements Review}

\author{Risha Parveen\inst{1} \and
Najmeh Akbari\inst{1} \and
Farnaz Delirie\inst{1} \and
Zheying Zhang\inst{1} \and
François Christophe\inst{2} \and
Jussi Rasku\inst{1} \and
Pekka Abrahamsson\inst{1}
}
\authorrunning{R. Parveen et al.}
\institute{
Tampere University, Tampere, Finland\
\email{{risha.parveen,najmeh.akbari,farnaz.delirie,zheying.zhang,
\\jussi.rasku,pekka.abrahamsson}@tuni.fi}
\and
University of Jyväskylä, Jyväskylä, Finland\
\email{francois.m.christophe@jyu.fi}
}

\maketitle

\begin{abstract}

Missing or underrepresented non-functional requirements (NFR) categories leave
little direct evidence in a requirements set, making them difficult to identify during requirements 
review. In this study, we frame NFR coverage-gap review as a project-level ranking task: given a target requirements specification, the goal is to rank the
11 PROMISE NFR categories for analyst inspection. We compare seven
statistical, embedding-based, and large language model (LLM)-based
configurations that differ in both ranking mechanism and evidence source, including 
target requirement statements, peer requirements, their semantic representation, and peer category statistics. 
We evaluate the configurations using controlled gap
injection with leave-one-project-out evaluation on PROMISE, PROMISE+,
PURE-Transport, and PURE-Business. Four configurations that use peer category
statistics are effectively tied under complete removal (0.875--0.889 mean
AUROC), despite different ranking mechanisms, and supplying those statistics to a
direct LLM ranker produces the largest evidence gain. All
configurations weaken when a category is present but underrepresented, and none
is best for every NFR category. For count-changing category-level gaps, 
structured peer evidence matters more than ranker
sophistication. The findings suggest that the rankings should guide analyst inspection rather than
serve as automatic judgements of requirements completeness.

\end{abstract}

\section{Introduction}

Following Glinz~\cite{glinz2007non}, NFRs comprise quality requirements and
constraints. Quality requirements describe attributes such as security,
performance, and usability, while constraints restrict how a system may be
built through mandated technologies, development rules, or external obligations.
This study focuses on quality attributes and uses the 11 PROMISE NFR
categories~\cite{cleland2007automated}, which also include legal and
operational concerns.

Assessing the coverage of these categories is difficult in requirements
review. A missing or underspecified category leaves little evidence
in the specification: reviewers can directly inspect what is documented, but identifying what has been omitted is more challenging. Under-documented quality concerns are nonetheless
consequential and have been studied as a source of technical
debt~\cite{scott2025underdocumented}. Detecting such gaps therefore requires
a reference beyond the target requirements.

Peer projects can provide this reference. Applications in related domains
often share recurring quality concerns~\cite{ramos2018nfr,ramos2025datadriven}:
for example, banking systems may repeatedly specify security and legal concerns, while
transport systems repeatedly specify availability and fault tolerance. We
therefore assume access to labelled requirements from a group of
comparable projects, which we call 
the peer
corpus. When an NFR category is commonly specified in the peer corpus but weakly
specified or absent in the target project, it becomes a candidate for analyst review.
The task studied here is to rank the 11 categories so that the most plausible
gaps appear first. The ranking supports analyst inspection; it does not
determine that additional requirements are necessary. The peer corpora in this study vary in domain scope: PROMISE and PROMISE+ provide mixed-domain peer corpus, whereas the PURE subsets provide smaller domain-scoped peer corpus. 

We compare seven ranking methods from three families: statistical,
embedding-based, and LLM-based. The methods differ along two dimensions: the mechanism used to produce a ranking and the evidence available to that mechanism. 
The evidence ranges from target requirement text alone to peer corpus, their category statistics, and semantic representation of peer requirements.
Because some methods share a
mechanism but differ in evidence, the study examines not only which
configuration rank well, but also where the ranking signal comes from.
The methods are evaluated on four datasets using controlled removal to create
known missing and underrepresented NFR categories. We address the following
research questions:

\begin{itemize}
    \item \textbf{RQ1:} How well do different ranking mechanisms identify
    missing or underrepresented NFR categories?
    \item \textbf{RQ2:} Which type of evidence provides the most useful signal
    for NFR coverage-gap ranking?
\end{itemize}


The remainder of this paper is organized as follows. Section~\ref{sec:background} reviews related work. Section~\ref{sec:study-design} presents the study design.
Section~\ref{sec:results} reports the results, Section~\ref{sec:discussion} discusses the findings and threats
to validity, and Section~\ref{sec:conclusion} concludes.

\section{Background and Related Work}
\label{sec:background}

Requirements review is an established quality gate between specification and
implementation. It ranges from informal walkthroughs to formal meetings with
defined roles for identifying anomalies and omissions in the requirements
document~\cite{basili1996empirical}. Completeness is one of the qualities
assessed during these reviews and is the focus of this study.

\begin{figure}[h]
    \centering
    \includegraphics[width=\linewidth]{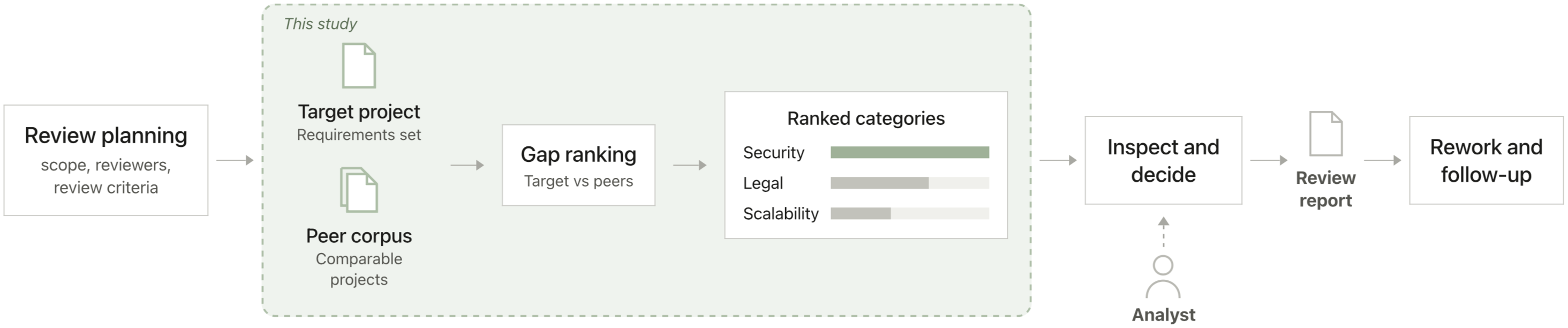}
    \caption{Peer-based coverage-gap ranking within a requirements review. This
    study covers the shaded steps: constructing the peer corpus, ranking the NFR
    categories, and reporting the coverage evidence behind each rank. Inspection,
    acceptance or rejection, and the review report remain analyst activities and
    are outside the evaluated scope.}
    \label{fig:review}
\end{figure}

Most review activities examine requirements that are already documented, for
example, whether a statement is unambiguous, verifiable, or internally
consistent. Coverage gaps are harder to assess because an absent or weakly
represented NFR category leaves little direct evidence in the target
specification. Generic quality checklists can suggest possible concerns, but
they do not show which concerns are expected for a specific project or how much
coverage is typical. We therefore add peer-based coverage-gap ranking to the
review process. It ranks NFR categories and provides supporting evidence, while
inspection, acceptance or rejection, and reporting remain analyst activities,
as shown in Fig.~\ref{fig:review}.

A \emph{peer corpus} is a set of comparable earlier projects used as a
reference for a target project, and \emph{peer requirements} are the individual
statements in that corpus. \emph{Peer statistics} are category-level counts and
proportions computed from the corpus, while \emph{peer evidence} includes peer
requirements, their embeddings, and these statistics. An \emph{NFR coverage
gap} is a category that is absent or underrepresented in the target project
relative to its peer corpus.

Table~\ref{tab:related-work} summarises related work and what remains open for
project-level NFR coverage-gap ranking. Statement-level checking and NFR
classification operate on requirements that are already present, so they offer
limited support when an entire category is undocumented. Reference-based
checking compares requirements with elicitation notes or domain
models~\cite{ferrari2014measuring,arora2019empirical,zhao2024usefulness}, but
suitable project-specific references may not be available. Language-model
approaches use term suggestions, coverage judgements, or generated
repairs~\cite{luitel2024improving,preda2024supporting,sheng2025recompgpt}, but
mainly reflect general expectations rather than whether a category is unusually
weak for a particular project. Requirements recommendation is closer to our
work because it uses historical projects, but it usually recommends candidate
statements rather than ranking categories or measuring category-level deficits.

\begin{table}[t]
\caption{Related work and remaining gap for project-level NFR
coverage-gap ranking.}
\label{tab:related-work}
\centering
\scriptsize

\setlength{\tabcolsep}{4pt}
\renewcommand{\arraystretch}{1.08}

\begin{tabularx}{\linewidth}{
|>{\hsize=0.75\hsize\raggedright\arraybackslash}X
|>{\hsize=1.20\hsize\raggedright\arraybackslash}X
|>{\hsize=1.05\hsize\raggedright\arraybackslash}X|
}
\hline

\textbf{Approach}
& \textbf{Typical evidence or output}
& \textbf{Remaining gap for this study} \\
\hline

Statement-level checking%
~\cite{eckhardt2016performance,liu2021conditional,lee2004missing}
& Checks written requirements for missing elements, semantic cues, or
viewpoint inconsistencies.
& Cannot flag an NFR category with no statement to inspect. \\
\hline

Reference-based checking%
~\cite{ferrari2014measuring,arora2019empirical,zhao2024usefulness}
& Compares requirements with elicitation material, domain models, or similar
references.
& Requires a suitable project-specific reference. \\
\hline

Language-model assistance%
~\cite{luitel2024improving,preda2024supporting,sheng2025recompgpt}
& Uses model predictions, coverage judgements, or generated repairs.
& Reflects general expectations rather than peer-relative coverage. \\
\hline

Requirements recommendation%
~\cite{ramos2018nfr,ramos2025datadriven}
& Uses historical project data to recommend candidate NFRs.
& Recommends statements rather than ranking category-level deficits. \\
\hline

NFR classification%
~\cite{cleland2007automated,hey2020norbert,alhoshan2023zero}
& Assigns individual requirements to NFR categories.
& Adds NFR labels only to individual requirements. \\
\hline

Peer-based gap ranking%
~\cite{yang2026detecting}
& Uses peer statistics and semantic representations to rank NFR categories.
& Requires evaluation and comparison; bounded to gaps that change category
counts. \\
\hline

\end{tabularx}
\end{table}

Three limitations motivate this study. First, approaches that inspect or classify existing requirement statements cannot flag a category that is not documented. Second, reference-based checking depends on the availability of suitable project-specific references. Third, language model judgement reflects general expectations rather than a project-specific comparison. 

Peer projects address these limitations by providing empirical evidence about recurring quality concerns for comparable projects.
What remains unclear 
is which form of peer evidence carries the signal, and
whether it matters more than the ranking mechanism applied to it. We therefore
frame NFR coverage-gap identification as a project-level category-ranking task
and compare target-only, statistical, semantic, and LLM-based evidence under one
evaluation. The contribution is an empirical understanding of evidence for NFR
coverage-gap ranking, rather than a new requirement classifier, requirement
generator, or automatic completeness judgement.

\section{Study Design}
\label{sec:study-design}

This study compares seven methods for ranking NFR coverage gaps. We use
the 11 NFR categories from the PROMISE classifier~\cite{cleland2007automated}:
availability, fault tolerance, legal, look and feel, maintainability,
operational, performance, portability, scalability, security, and usability.
For each target project, every configuration ranks these categories from most
to least likely to be missing or underrepresented. The ranking directs analyst
attention; the analyst determines whether the candidate gap is relevant to the project.

\begin{table}[h]
\centering
\caption{Ranking methods evaluated in the study.}
\label{tab:methods}

{\scriptsize
\setlength{\tabcolsep}{2pt}
\renewcommand{\arraystretch}{1.1}

\begin{tabular}{
|p{0.17\linewidth}|
p{0.17\linewidth}|
p{0.28\linewidth}|
p{0.31\linewidth}|
}
\hline
\textbf{Configuration}
& \textbf{Family}
& \textbf{Mechanism}
& \textbf{Evidence used for ranking} \\
\hline
Count deficit
& Statistical
& Standardised count-deficit scoring
& Target and peer category counts \\
\hline
Ratio deficit
& Statistical
& Standardised proportion-deficit scoring
& Target and peer category proportions \\
\hline
kNN retrieval
& Embedding-based
& Local nearest-neighbour distance scoring
& Qwen3 embeddings of target and peer requirements \\
\hline
GEOGAP~\cite{yang2026detecting}
& Embedding-based
& Category-aware geometric coverage scoring
& Qwen3 embeddings and category-population statistics \\
\hline
Zero-shot LLM
& LLM-based
& Direct LLM ranking
& Target requirement text \\
\hline
Statistics-aware LLM
& LLM-based
& Direct LLM ranking
& Target text and structured target--peer statistics \\
\hline
Stat-guided LLM reranker
& LLM-based
& Ratio-deficit screening followed by LLM reranking
& Target text and structured target--peer statistics \\
\hline
\end{tabular}
}
\end{table}

\subsection{Ranking Methods}
\label{sec:ranking-methods}

Table~\ref{tab:methods} summarises the seven ranking methods evaluated in the study. They are grouped by mechanism as statistical, embedding-based, and
LLM-based. The evidence used for ranking varies across and within these
families. The comparison is therefore about both ranking mechanisms and evidence sources, rather than a simple contest among model families. 

\textit{Statistical methods.}
The statistical methods compare 
how NFR category $c$ is represented in the target project $P_t$ with 
how the same category is represented in the peer projects $P_p$. Count deficit measures the standardised shortfall in the number of target requirements assigned to $c$, while Ratio deficit measures the standardised shortfall in the proportion of target requirements assigned to $c$, thereby accounting for specification size. The scores are:

\begin{equation}
\scriptsize
S_{\mathrm{count}}(c)
=
\frac{
\text{mean count of } c \text{ across } P_p
-
\text{count of } c \text{ in } P_t
}{
\text{standard deviation of counts of } c \text{ across } P_p
},
\label{eq:count-deficit}
\end{equation}

\begin{equation}
\scriptsize
S_{\mathrm{ratio}}(c)
=
\frac{
\text{mean proportion of } c \text{ across } P_p
-
\text{proportion of } c \text{ in } P_t
}{
\text{standard deviation of proportions of } c \text{ across } P_p
}.
\label{eq:ratio-deficit}
\end{equation}

Higher scores indicate lower target coverage, particularly when peer coverage is consistent. Because both evaluation protocols reduce category counts and proportions, these methods serve as protocol-aligned baselines for assessing how much of the ranking signal can be recovered from coverage statistics alone. The embedding- and LLM-based methods test whether richer evidence improves on these baselines.



\textit{Embedding-based methods.}
kNN retrieval measures local semantic coverage using nearest-neighbour
distance~\cite{cover1967nearest}. For each peer requirement, it finds the
closest target requirement and aggregates how unusual these distances are by
NFR category. A high score indicates that the peer requirements of a category
are poorly represented in the target. GEOGAP~\cite{yang2026detecting} extends
this approach by combining general semantic matching with same-category
matching and information about whether the target contains fewer requirements
for the category than its peers. Both methods use
Qwen3-Embedding-0.6B~\cite{zhang2025qwen3}, so their comparison examines
whether a local distance signal is sufficient or whether combining semantic
and category-population evidence is more useful.


\textit{LLM-based methods.}
Following previous applications of language models to requirements completeness
and coverage review~\cite{luitel2024improving,preda2024supporting}, we evaluate
one target-only and two peer-informed methods. The Zero-shot LLM receives
the target requirements and definitions of the 11 NFR categories and directly
ranks all categories without examples, statistics, or peer information. The
Statistics-aware LLM additionally receives target category counts and
proportions and the corresponding peer means and standard deviations, allowing
their comparison to measure the effect of adding peer statistics while keeping
direct LLM ranking fixed. The Stat-guided LLM reranker first selects the seven
highest-scoring categories using Ratio deficit and reranks them using the target
requirements and peer statistics, while the remaining four retain their
Ratio-deficit order. Seven candidates are used because preliminary screening
retained $94$--$100\%$ of the removed categories, following the
retrieve-then-rerank principle~\cite{meng2024ranked}. Comparing the reranker with the Statistics-aware LLM contrasts direct ranking with statistical screening followed by contextual reranking. All three methods use GPT-5.1 at temperature $0.0$ under consistent prompting and output settings.

\subsection{Datasets and Peer Construction}
\label{sec:datasets}

We evaluate the methods on PROMISE~\cite{cleland2007automated},
PROMISE+~\cite{silva2024promise+}, PURE-Transport, and PURE-Business.
Functional requirements are removed, and malformed, empty, or unlabelled
entries are excluded. All datasets are mapped to the 11 PROMISE NFR
categories. For each eligible target, all remaining projects in the same
dataset form its peer corpus, and the peer-based evidence is recomputed.
Table~\ref{tab:datasets} summarises the dataset sizes and target eligibility;
the eligibility criteria are defined in Section~\ref{sec:evaluation}.

\begin{table}[h]
\centering
\caption{Dataset size and target eligibility. Category size is the number of
requirements per non-empty project--category pair, reported as median
(maximum).}
\label{tab:datasets}

{\scriptsize
\setlength{\tabcolsep}{4pt}
\renewcommand{\arraystretch}{1.1}

\begin{tabular}{|l|l|r|r|r|r|r|}
\hline
\textbf{Dataset}
& \textbf{Domain}
& \textbf{\#Projects}
& \textbf{\#NFRs}
& \textbf{Category size}
& \multicolumn{2}{c|}{\textbf{Eligible targets}} \\
\cline{6-7}
& & & & \textbf{Median (max.)} & \textbf{Fixed} & \textbf{Underrep.} \\
\hline
PROMISE
& Mixed
& 15
& 370
& 3 (17)
& 15
& 10 \\
\hline
PROMISE+
& Mixed
& 79
& 1,419
& 2 (153)
& 59
& 26 \\
\hline
PURE-Transport
& Transport
& 5
& 133
& 2 (14)
& 5
& 4 \\
\hline
PURE-Business
& Business
& 9
& 504
& 5 (42)
& 9
& 8 \\
\hline
\end{tabular}
}
\end{table}

PROMISE is a small mixed-domain benchmark, while PROMISE+ is a larger
extension that contains PROMISE. Their comparison is therefore treated as a
sensitivity check rather than an independent replication. PURE-Transport and
PURE-Business were derived from the PURE collection~\cite{ferrari2017pure} and
kept separate to preserve domain-specific peer corpora. Candidate NFR statements
were extracted with LLM assistance, labelled using the prompt-based
classification approach of Binkhonain and Alfayaz~\cite{binkhonain2025prompts},
and reviewed by the authors. We use these subsets for external validation while
recognising that their extraction and labelling were author-supervised rather
than independently annotated.

\subsection{Evaluation Protocols}
\label{sec:evaluation}

\begin{figure}[h]
    \centering
    \includegraphics[width=\linewidth]{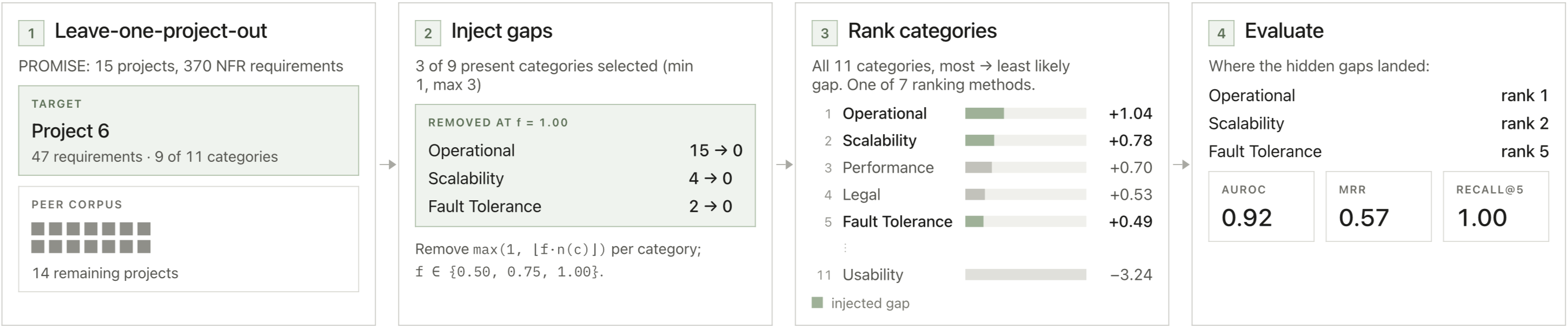}
    \caption{Evaluation workflow for creating known NFR coverage gaps and
    evaluating the resulting category rankings.}
    \label{fig:evaluation-workflow}
\end{figure}

The datasets show which NFR categories are present, but not which categories
should be present. We therefore create known gaps by removing requirements from
selected categories in each target project, following the evaluation approach
of Arora et al.~\cite{arora2019empirical}. As shown in
Fig.~\ref{fig:evaluation-workflow}, each ranking configuration receives the
modified target, ranks all 11 categories, and is evaluated against the injected
categories.
We use two protocols to represent different forms of coverage gap. The
fixed-fraction protocol partially or completely removes a category, whereas the
underrepresentation protocol keeps the category present but reduces it below
typical peer coverage. Both protocols evaluate count-changing category-level
gaps. 

\textit{Fixed-fraction protocol.}
We retain projects with at least five requirements across at least two NFR
categories. Categories containing at least three requirements are eligible for
removal; projects with no such category are excluded. For each target,
approximately one third of its eligible categories are selected, with a minimum
of one and a maximum of three. From each selected category, we remove
$\max(1,\lfloor f\,n_p(c)\rfloor)$ requirements, where $n_p(c)$ is the
original category count and $f \in \{0.50,0.75,1.00\}$. A seed derived from
the project identifier ensures that every configuration receives the same
injected gaps. Complete removal, $f=1.00$, is the main missing-category
condition; $f=0.50$ and $f=0.75$ represent weaker removal, which does not
guarantee that the remaining coverage is low relative to the peers.

\textit{Underrepresentation protocol.}
This protocol reduces a category below its typical peer coverage while
retaining at least one requirement. The peer coverage floor $\tau_p(c)$ is the
lower quartile of $\{r_{p'}(c) : p' \in \mathrm{peers}(p)\}$, where
$r_{p'}(c)$ is the proportion of peer project $p'$ assigned to category $c$.
If the quartile is zero, it is recalculated using only peers that contain the
category, provided that at least four such peers are available. A category is
eligible when it contains at least three requirements, starts at or above the
floor, and can fall below it while retaining one requirement. Up to three
eligible categories are selected per target, and requirements are removed until
each category proportion falls below $\tau_p(c)$. Because many small categories
cannot meet these conditions without being removed completely, this protocol
yields fewer eligible targets than the fixed-fraction protocol
(Table~\ref{tab:datasets}).

\textit{Evaluation metrics.}
Injected categories are treated as relevant, and metrics are computed per
target and macro-averaged within each dataset and condition.
AUROC~\cite{hanley1982meaning} measures how often an injected category is
ranked above a non-injected one, with $1.0$ indicating perfect ordering;
MRR~\cite{voorhees1999trec} averages the reciprocal ranks of the injected
categories; and Recall@5~\cite{manning2008introduction} is the proportion of
injected categories placed in the top five. For $K=11$ categories, the chance
levels under a uniformly random ranking are
\begin{equation}
\scriptsize
C_{\mathrm{AUROC}}=0.50,\qquad
C_{\mathrm{MRR}}=\frac{H_{11}}{11}\approx0.27,\qquad
C_{\mathrm{R@5}}=\frac{5}{11}\approx0.45,
\label{eq}
\end{equation}
where $H_{11}=\sum_{r=1}^{11}1/r$. These constants serve as chance-level
references in the analysis and figures.

\subsection{Analysis Approach}
\label{sec:analysis-plan}
The methods differ in both ranking mechanism and evidence, so we compare
them at three levels: individual configuration, targeted within-family contrasts, and peer corpus size and composition sensitivity analysis. AUROC is
the primary measure, with MRR and Recall@5 reported alongside it. Results are
interpreted against the analytic in Eq.~\ref{eq} and the protocol-aligned statistical baselines.
For RQ1, we compare all seven methods under partial removal, complete
removal, and underrepresentation. We examine whether each performs above
chance, whether the embedding- and LLM-based methods
match or improve on
the statistical baselines, and whether their relative performance changes with
removal severity, gap type, dataset, or NFR category. At the category level, we
also use Spearman's $\rho$ across category--dataset combinations to relate mean
AUROC to mean peer coverage and assess whether recoverability tracks how
commonly a category is documented in the peer corpus.

For RQ2, we report four targeted within-family contrasts as differences in mean
AUROC between configuration pairs (Table~\ref{tab:contrasts}). These contrasts
are diagnostic rather than a factorial ablation: only the
Zero-shot--Statistics-aware comparison holds the ranking mechanism fixed while
changing the evidence; the others also change how evidence is represented,
organised, or sequenced. We further compare PROMISE with PROMISE+,
PURE-Transport with PURE-Business, and the mixed-domain PROMISE datasets with
the domain-specific PURE datasets to examine the amount and domain composition
of peer evidence. Because PROMISE is contained in PROMISE+ and the dataset
pairs vary in size and domain together, these comparisons are treated as
sensitivity analyses rather than controlled tests. All code, data, and result files are publicly available.\footnote{\url{https://github.com/risha-parveen/nfr_gap_ranking_study}}

\section{Results}
\label{sec:results}

The evaluation covers 2,426 NFR statements from 108 target projects across the four datasets. Eligibility filtering leaves 88 targets for the fixed-fraction
protocol and 48 for underrepresentation. The seven methods produce
1,848 and 336 rankings, so 2,184 in total, 936 of which use an LLM. Each
ranking orders all 11 categories. Section~\ref{sec:rq1} compares the
methods as complete ranking approaches (RQ1) and
Section~\ref{sec:rq2} traces where the ranking signal comes from (RQ2).

\subsection{RQ1: How Well Are Coverage Gaps Identified?}
\label{sec:rq1}

RQ1 compares the seven methods across removal severity, gap type, and
NFR category. AUROC is the main measure. MRR and Recall@5 are reported in Table~\ref{tab:drop}

\begin{figure}[h]
    \centering
    \includegraphics[width=\linewidth]{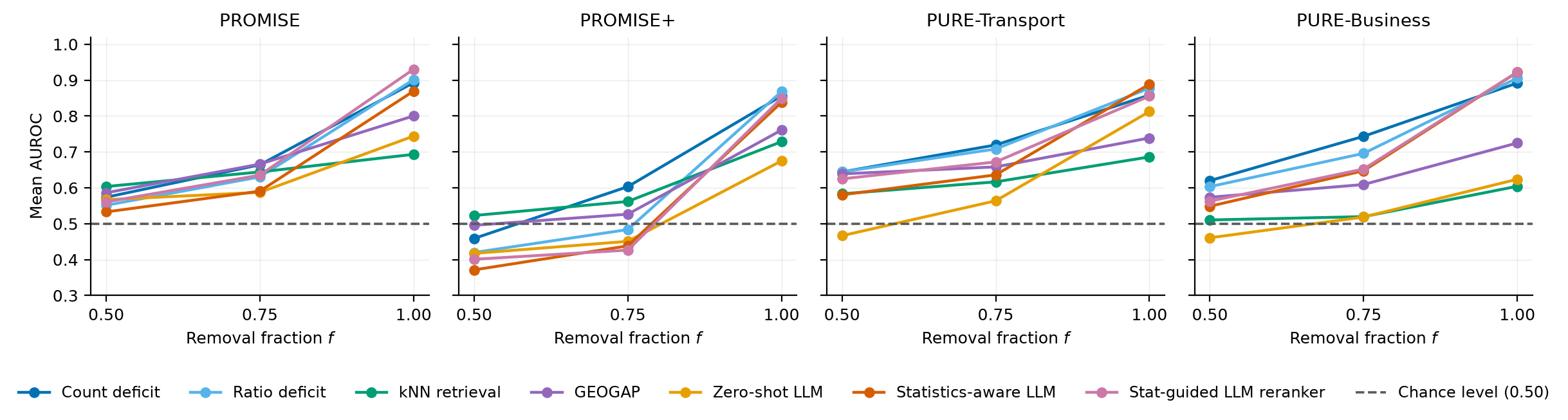}
    \caption{Mean AUROC as a function of the removal fraction $f$, per dataset.
    The right edge ($f=1.00$) is the complete-removal condition reported in
    Table~\ref{tab:drop}.}
    \label{fig:fraction}
\end{figure}

\subsubsection{Removal Severity and Gap Type}

Under complete removal, four methods are effectively tied at
0.875--0.889 mean AUROC: Count deficit, Ratio deficit, the Statistics-aware
LLM, and the Stat-guided LLM reranker. Their spread of 0.014 is smaller than
the variation across datasets, so we do not identify a single best
configuration. GEOGAP, the Zero-shot LLM, and kNN retrieval are clearly
weaker. The four leading methods all use peer category statistics,
either directly or as LLM input, and they also form the leading group on MRR
and Recall@5.

At weaker fixed-fraction removal, the methods are only slightly
separated, several fall below chance on PROMISE+ at $f=0.50$, and clear
differences emerge only as $f$ approaches 1.00. kNN retrieval changes least
with removal severity: it is strongest on both PROMISE datasets at $f=0.50$
but weakest under complete removal. This instability occurs because removing a
fixed proportion does not ensure that the resulting category coverage is low
relative to the peers. The underrepresentation protocol addresses this by
reducing coverage below the peer floor while keeping the category present.
Although it often removes fewer requirements than $f=0.50$, it produces a
stronger signal in every dataset. Detectability therefore depends mainly on
the final coverage relative to the peers, rather than on the number of removed
requirements.

\begin{table}[h]
\centering
\scriptsize
\caption{Mean AUROC, MRR, and Recall@5 under complete removal
($f=1.00$) and underrepresentation, macro-averaged across the four
datasets. Rows are ordered by the AUROC decrease. Chance levels are
0.50, 0.27, and 0.45.}
\label{tab:drop}

\setlength{\tabcolsep}{3.5pt}
\renewcommand{\arraystretch}{1.08}

\begin{tabularx}{\linewidth}{
|>{\raggedright\arraybackslash}X
|c|c|c
|c|c|c|
}
\hline
\multicolumn{1}{|c|}{}
& \multicolumn{3}{c|}{\textbf{Complete removal ($f=1.00$)}}
& \multicolumn{3}{c|}{\textbf{Underrepresentation}} \\
\hline

\textbf{Configuration}
& \textbf{AUROC}
& \textbf{MRR}
& \textbf{R@5}
& \textbf{AUROC}
& \textbf{MRR}
& \textbf{R@5} \\
\hline

kNN retrieval
& 0.678 & 0.417 & 0.687
& 0.609 & 0.324 & 0.611 \\
\hline

GEOGAP
& 0.756 & 0.506 & 0.776
& 0.653 & 0.410 & 0.632 \\
\hline

Count deficit
& 0.875 & 0.594 & 0.893
& 0.732 & 0.361 & 0.772 \\
\hline

Ratio deficit
& 0.888 & 0.618 & 0.913
& 0.703 & 0.367 & 0.667 \\
\hline

Zero-shot LLM
& 0.714 & 0.407 & 0.761
& 0.487 & 0.247 & 0.438 \\
\hline

Statistics-aware LLM
& 0.880 & 0.570 & 0.928
& 0.639 & 0.283 & 0.650 \\
\hline

Stat-guided LLM reranker
& 0.889 & 0.609 & 0.909
& 0.619 & 0.263 & 0.641 \\
\hline
\end{tabularx}
\end{table}

All methods weaken under underrepresentation, but their ordering
changes. Count deficit becomes strongest at 0.73 because it degrades less than
the alternatives, not because its performance improves. In general, the
methods that benefit most from complete absence lose most when the
category remains present. The Zero-shot LLM falls to chance overall, showing
that target text alone provides no reliable indication that a present category
is unusually thin. The metric-specific results add one qualification: GEOGAP
has the highest MRR despite ranking fourth on AUROC, suggesting that its
embedding evidence often places a gap near the top even when its full-list
ordering is weaker. Count deficit leads on both AUROC and Recall@5.

\subsubsection{Category-Level Recoverability}

\begin{figure}[t]
    \centering
    \includegraphics[width=0.85\linewidth]{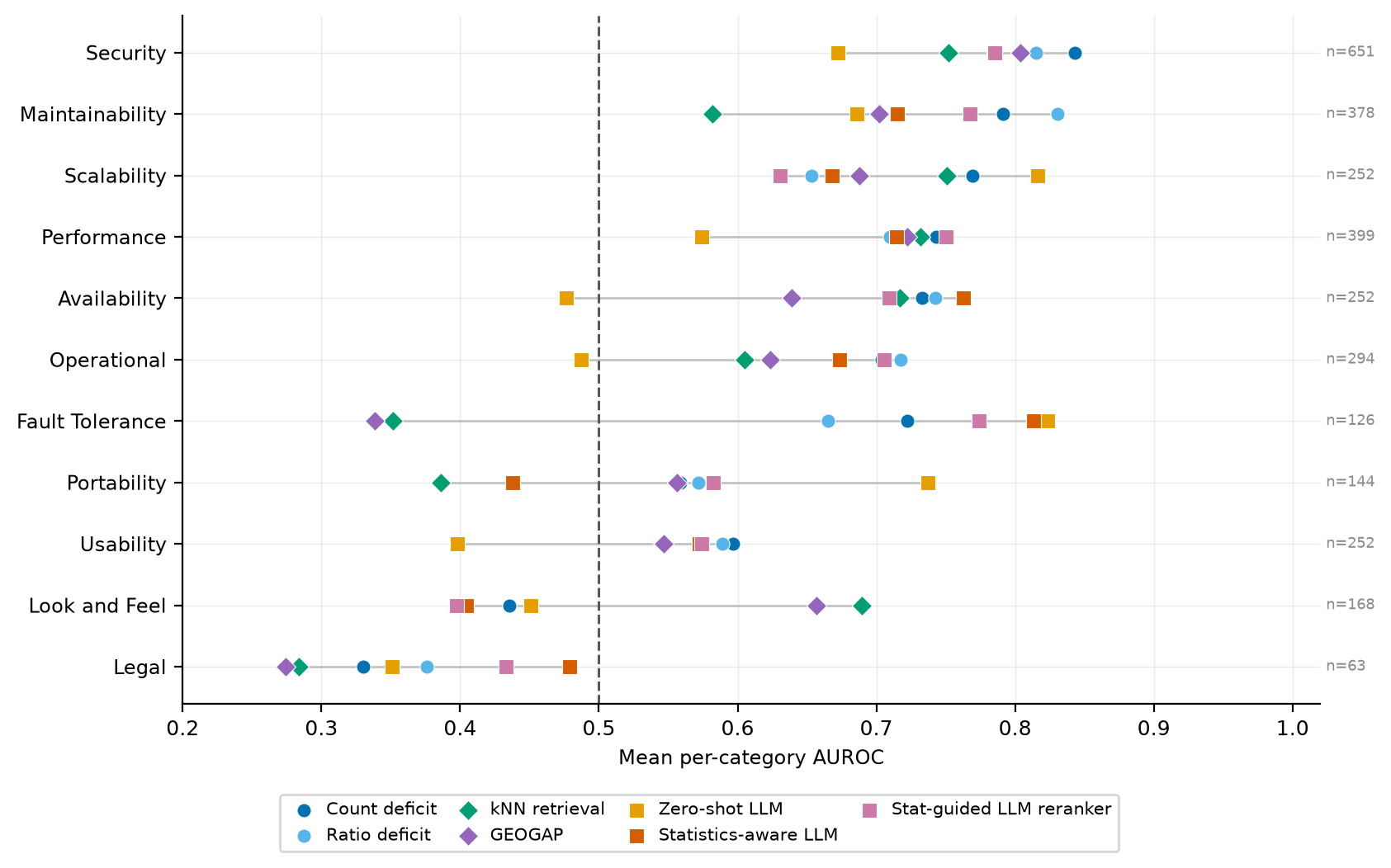}
    \caption{Mean AUROC by NFR category under complete removal, averaged across
    datasets, with one marker per configuration. $n$ is the number of
    requirements in the category across all datasets.}
    \label{fig:category}
\end{figure}

Recoverability varies more across NFR categories than across methods
(Fig.~\ref{fig:category}). Security and Maintainability are the most
recoverable, while Legal is the least. Across the 38 non-empty category--dataset
combinations, mean AUROC is positively correlated with peer coverage
($\rho=0.59$), indicating that frequently documented categories provide a
more stable reference than rare or project-specific concerns.
The methods also differ by category. Look and Feel is above chance only
for kNN retrieval and GEOGAP, whereas Fault Tolerance is recovered well by the
statistical and LLM-based methods but poorly by both embedding-based
ones. No configuration performs best across all categories, suggesting that
the evidence sources capture partly complementary signals.

\subsection{RQ2: Which Evidence Sources Are Most Useful?}
\label{sec:rq2}

\begin{figure}[h]
     \centering
     \includegraphics[width=\linewidth]
     {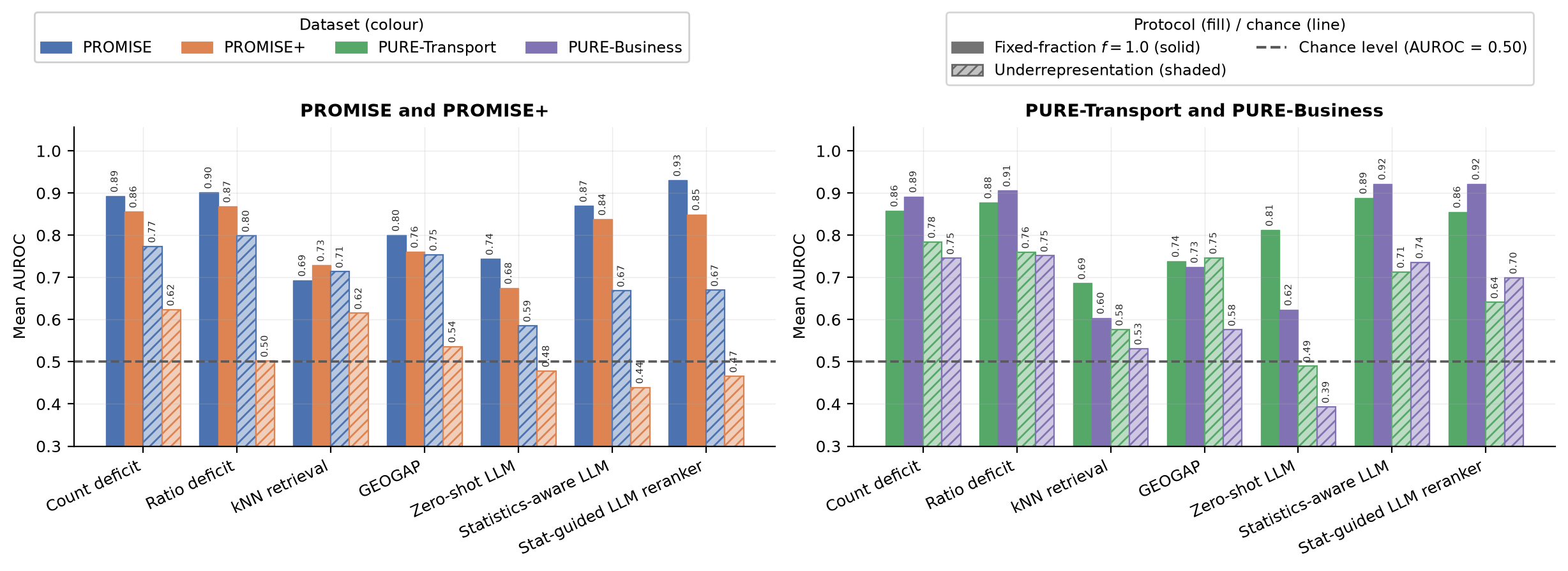}
     \caption{Mean AUROC per configuration under complete removal and
     underrepresentation across the four datasets. Dashed lines indicate the
     analytic chance level.}
     \label{fig:grid}
\end{figure}

RQ2 examines which forms of evidence improve coverage-gap ranking. We compare
four pairs of related methods and then examine how performance changes
across the datasets. Only the Zero-shot and Statistics-aware LLM comparison
changes the evidence while keeping the ranking mechanism fixed. The other
comparisons also change how the evidence is represented or used, so they do not
isolate a single factor. Fig.~\ref{fig:grid} reports the per-dataset results.

\subsubsection{Form and Use of Evidence}

Table~\ref{tab:contrasts} shows that adding peer category statistics produces
the largest improvement. The other changes have smaller and less consistent
effects.

\begin{table}[h]
\centering
\scriptsize
\caption{Targeted within-family contrasts. Values are differences in mean
AUROC across the four datasets.}
\label{tab:contrasts}

\setlength{\tabcolsep}{4pt}
\renewcommand{\arraystretch}{1.08}

\begin{tabularx}{\linewidth}{
|>{\raggedright\arraybackslash}p{0.29\linewidth}
|>{\raggedright\arraybackslash}X
|c
|c|
}
\hline
\textbf{Contrast}
& \textbf{Difference examined}
& \textbf{$f=1.00$}
& \textbf{Underrep.} \\
\hline

Zero-shot \(\rightarrow\) Statistics-aware LLM
& Target text only \(\rightarrow\) target text and peer statistics
& \(+0.166\)
& \(+0.152\) \\
\hline

kNN retrieval \(\rightarrow\) GEOGAP
& Local semantic distance \(\rightarrow\) category-aware semantic and
population evidence
& \(+0.078\)
& \(+0.044\) \\
\hline

Count \(\rightarrow\) Ratio deficit
& Absolute count shortfall \(\rightarrow\) size-normalised proportion
shortfall
& \(+0.013\)
& \(-0.029\) \\
\hline

Statistics-aware \(\rightarrow\) Reranker
& Direct LLM ranking \(\rightarrow\) statistical screening and LLM reranking
& \(+0.009\)
& \(-0.020\) \\
\hline
\end{tabularx}
\end{table}

Adding peer category statistics improves the direct LLM ranker by 0.166 mean
AUROC under complete removal and 0.152 under underrepresentation. Under
complete removal, both LLM methods are already above chance. Under
underrepresentation, however, the Zero-shot LLM is at or below chance on every
dataset, while the Statistics-aware LLM remains clearly above chance on both
PURE datasets. Target text can show whether a category is discussed, but it
cannot show whether that coverage is unusually low. Peer statistics provide
the reference needed to make this comparison.

GEOGAP performs better than kNN retrieval on average, but the comparison
changes two aspects at once. GEOGAP groups semantic evidence by category and
also includes category-population information. Its improvement therefore
cannot be attributed to either change alone, and the result varies across
datasets. GEOGAP gains most on PROMISE under complete removal, whereas kNN
retrieval performs better on PROMISE+ under underrepresentation.

Count and Ratio deficit perform similarly overall. Ratio deficit is slightly
stronger under complete removal but weaker under underrepresentation, mainly
because it falls to chance on PROMISE+. Ratio deficit becomes less useful when
peer projects differ widely in the expected proportion of a category, since
the large variation reduces the resulting deficit score. Statistical screening
before LLM reranking also gives no average accuracy advantage over direct
Statistics-aware ranking. The reranker performs better on three of four
datasets under underrepresentation, but its main practical benefit is that the
LLM ranks fewer categories.

\subsubsection{Peer-set Size and Composition}
\label{sec:rq2-peer-sets}

The dataset comparisons vary peer-set size and domain at the same time, so they
cannot show the independent effect of either factor. PROMISE+ contains about
5.3 times more projects than PROMISE but also has a broader mixed-domain peer
corpus. PURE-Transport and PURE-Business differ in both size and application
domain.

Performance falls from PROMISE to PROMISE+ for six of seven methods
under complete removal and for all seven under underrepresentation. In
contrast, under complete removal, every configuration using aggregate peer
statistics improves from PURE-Transport to PURE-Business, while kNN retrieval
declines. A larger peer set therefore does not necessarily provide a stronger
reference. For statistical methods, agreement among peers about
expected category coverage appears more important than the number of projects.

The mixed-domain PROMISE datasets and the domain-specific PURE datasets show
the same pattern. Under underrepresentation, the statistical baselines average
0.673 on PROMISE and PROMISE+ and 0.760 on PURE-Transport and PURE-Business.
Embedding-based evidence behaves differently: kNN retrieval performs better on
the mixed-domain datasets, while GEOGAP changes little. Domain consistency
therefore appears more useful for estimating expected category coverage than
for local semantic matching. However, the datasets also differ in size,
category distribution, and the number of valid targets.
\section{Discussion}
\label{sec:discussion}

This study compared seven methods for ranking missing or
underrepresented NFR categories. The main finding is that, for count-changing
category-level gaps, what evidence the ranker is given matters more than how
sophisticated the ranker is. Under complete removal, the leading methods
perform about the same, and all of them use peer category statistics. Once a
category is clearly thin compared with the peers, a more complex ranker adds
little.
Underrepresentation is harder, because the category is still present in the
target project, and every configuration performs worse there. Even so, it gives
a clearer signal than removing half the requirements at $f=0.50$, although it
often removes fewer of them. At $f=0.50$ the methods are also close
together and their order is unstable. What decides whether a gap can be found
is therefore not how many requirements were removed, but whether the remaining
coverage ends up below the peer reference. The methods only separate
once that shortfall is large enough.


The LLM results support the same observation. A target-only LLM
can identify generally expected quality concerns, but target text does not
indicate whether a present category is underrepresented.
Adding peer statistics to the LLM ranker provides that
project-relative reference and produces the largest evidence gain in the study.
This implies that LLMs should not be treated as standalone completeness
reviewers for this task. Their usefulness depends on whether they are supplied
with structured evidence about the target and its peers. The remaining
contrasts are smaller. GEOGAP combines category-aware semantic aggregation with
population evidence, while statistical screening reduces the number of
categories ranked by the LLM but provides no clear accuracy advantage over
direct Statistics-aware ranking. The relative performance of Count and Ratio
deficit also depends on variation among the peers.

No single evidence source is best overall; they appear to capture different
things. No configuration ranks best for every NFR category, only the
embedding-based methods place Look and Feel above chance, and GEOGAP has
the highest MRR under underrepresentation even though its full ranking is
weaker. A larger peer corpus also does not guarantee better rankings: PROMISE+ has
more projects than PROMISE but is weaker for most methods, and the PURE
subsets suggest that peers agreeing on how much of a category is normal helps
more than peer count. Rankings are likewise more reliable for categories that
are consistently covered than for rare or project-specific ones such as Legal.

\subsection{Implications for Requirements Review}

When labelled requirements from comparable earlier projects are available,
Count deficit provides a transparent and inexpensive first check for category
absence or unusually low category population. It requires no LLM or embedding
model, but depends on reliable NFR labels and a suitable peer corpus.
Embedding-based evidence may complement this check for categories such as Look
and Feel, but does not replace the stronger statistical signal. For LLM-supported
review, target requirements alone are insufficient; structured peer evidence
and careful peer selection should be treated as part of the review design.

The ranking should support inspection rather than determine completeness. A
high-ranked category indicates a difference from the selected peers and should
prompt analyst review, which may confirm an omission, establish that the
category is irrelevant, or reveal noise in the peer corpus. A low rank does not
prove adequate coverage because mild shortfalls are less reliably detected and
category-level methods may miss quality levels, exceptions, subtopics, and
domain-specific constraints.

\subsection{Threats to Validity}
\label{sec:threats}

For construct validity, both protocols create gaps by reducing category counts,
which favours Count and Ratio deficit and limits the findings to count-changing
gaps. Gaps that leave category counts unchanged, such as missing conditions or
quality levels, are outside the scope of this study. Among the RQ2 comparisons,
only the Zero-shot and Statistics-aware LLMs keep the ranking mechanism fixed
while changing the evidence; the others also change how the evidence is
organised or used. For internal validity, each target uses one seeded injection,
and the LLM methods use one model and prompt setup, so different
injections or LLM settings may produce different results. For conclusion
validity, we report point estimates without significance testing. The four
leading methods under complete removal lie within 0.014 mean AUROC,
which is smaller than the variation across datasets, so we do not claim an
ordering among them. Findings about the reranker and peer-set composition are
based on four datasets and remain descriptive. For external validity, the
datasets are research benchmarks with smaller peer corpus than typical industrial
collections, and PROMISE is contained in PROMISE+. The 11 PROMISE categories
may also omit concerns relevant to modern systems. The PURE subsets use
LLM-assisted extraction and classification followed by author review, which may
introduce labelling errors or favour the LLM-based methods. In practice,
peer labels may be assigned automatically rather than curated, which could
weaken the peer statistics used by the leading methods.

\section{Conclusion}
\label{sec:conclusion}

This paper framed NFR coverage-gap detection as a project-level
category-ranking task and compared seven statistical, embedding-based, and
LLM-based methods. The results show that, for count-changing gaps,
structured peer coverage is more useful than target text alone and often matters
more than ranking complexity. Simple peer-relative comparisons therefore provide
a strong basis for prioritising analyst review. However, performance weakens when
a category remains present but underrepresented, and no configuration is best
for every NFR category. The rankings should therefore support inspection rather
than determine completeness.

Future work should evaluate count-preserving gaps, vary peer-set size and
similarity independently, separate GEOGAP's components, and test combinations of
complementary evidence. Evaluation with practising analysts, industrial
requirements, and automatically assigned NFR labels is also needed.

\section*{Acknowledgements}

This work was supported by Business Finland through the ANSE project and 6GSoft project
(8548/31/2022)

\bibliographystyle{splncs04}
\bibliography{refs}

\end{document}